\documentclass[aps,pra,twocolumn,groupedaddress,showkeys]{revtex4}

\usepackage{graphics}

\begin{document}


\title{Bipartite Yule Processes in Collections of Journal Papers}

\author{Steven A. Morris}
\email[]{steven.a.morris@okstate.edu}
\homepage[]{http://samorris.ceat.okstate.edu}
\affiliation{
Oklahoma State University\\
Electrical and Computer Engineering\\
Stillwater, OK 74078, USA }

\date{\today}

\begin{abstract}
Collections of journal papers, often referred to as 'citation
networks', can be modeled as a collection of coupled bipartite
networks which tend to exhibit linear growth and preferential
attachment as papers are added to the collection. Assuming primary
nodes in the first partition and secondary nodes in the second
partition, the basic bipartite Yule process assumes that as each
primary node is added to the network, it links to multiple secondary
nodes, and with probability, $\alpha$, each new link may connect to
a newly appearing secondary node. The number of links from a new
primary node follows some distribution that is a characteristic of
the specific network. Links to existing secondary nodes follow a
preferential attachment rule. With modifications to adapt to
specific networks, bipartite Yule processes simulate networks that
can be validated against actual networks using a wide variety of
network metrics. The application of bipartite Yule processes to the
simulation of paper-reference networks and paper-author networks is
demonstrated and simulation results are shown to mimic networks from
 actual collections of papers across several network metrics.
\end{abstract}


\maketitle

\section{Collections of papers as coupled bipartite networks}

As shown in Figure 1, a collection of journal papers constitutes a
series of coupled bipartite networks \cite{morris05}. As diagrammed
in Figure 1, a collection of papers contains 6 direct bipartite
networks: 1) papers to paper authors, 2) papers to references, 3)
papers to paper journals, 4) papers to terms, 5) references to
reference authors, and 6) references to reference journals.
Additionally, there are 15 indirect bipartite networks in
collections of papers as defined by the diagram. Examples of
interesting indirect networks are paper author to reference author
networks, and paper journal to reference journal networks, which can
be used for author co-citation analysis \cite{white81} and journal
co-citation analysis \cite{mccain91} respectively.

\begin{figure}
\resizebox{0.45\textwidth}{!}{%
\includegraphics{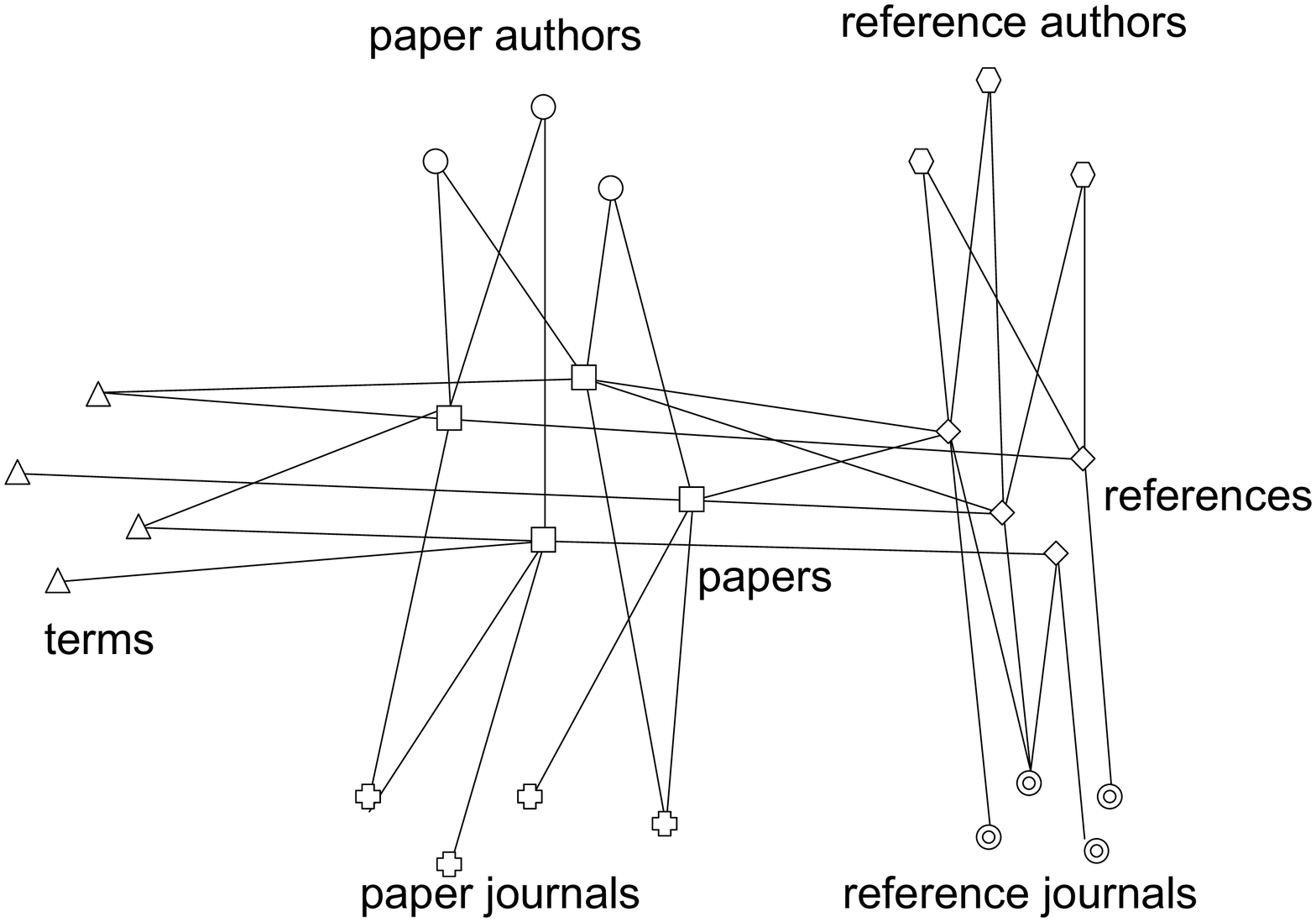}}%
\caption{Diagram showing a collection of papers as a series of
coupled bipartite networks.\label{coupled}}
\end{figure}

Modeling the growth of these bipartite networks helps characterize
the underlying processes driving a research specialty, such as
knowledge accretion, researcher productivity, or collaboration
processes. Bipartite growth models produce many network metrics,
allowing comprehensive validation of models against real collections
of papers.

\section{Basic bipartite Yule processes}

As originally proposed, Yule processes do not model networks, but
simply model the formation of power-laws of frequencies of items
\cite{albert02} \cite{price76} \cite{simon55}. For a bipartite Yule
process, assume a bipartite network where nodes fall into two
partitions: 1) primary nodes and 2) secondary nodes. Typically,
primary nodes are papers while secondary nodes are entities that are
associated with papers, such as authors, references,  journals, or
terms.

Figure 2 shows a diagram of a bipartite paper-reference network,
where the primary nodes are papers and the secondary nodes are
references, and papers are linked to references by citations.

\begin{figure}
\resizebox{0.35\textwidth}{!}{%
\includegraphics{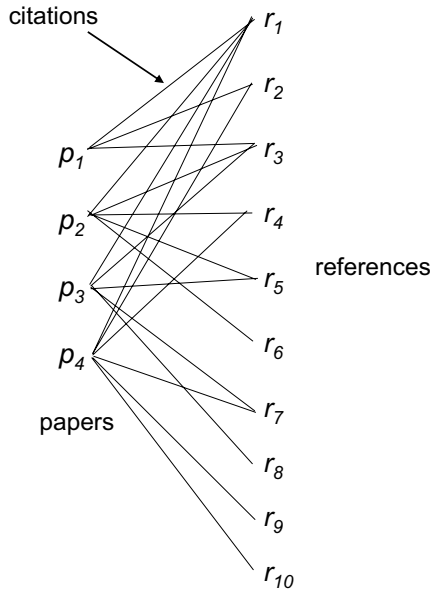}}%
\caption{Diagram showing a bipartite network of papers and the
references that they cite.\label{pr}}
\end{figure}

Figure 3 shows a diagram of a basic bipartite Yule process:

\begin{figure}
\resizebox{.45\textwidth}{!}{%
\includegraphics{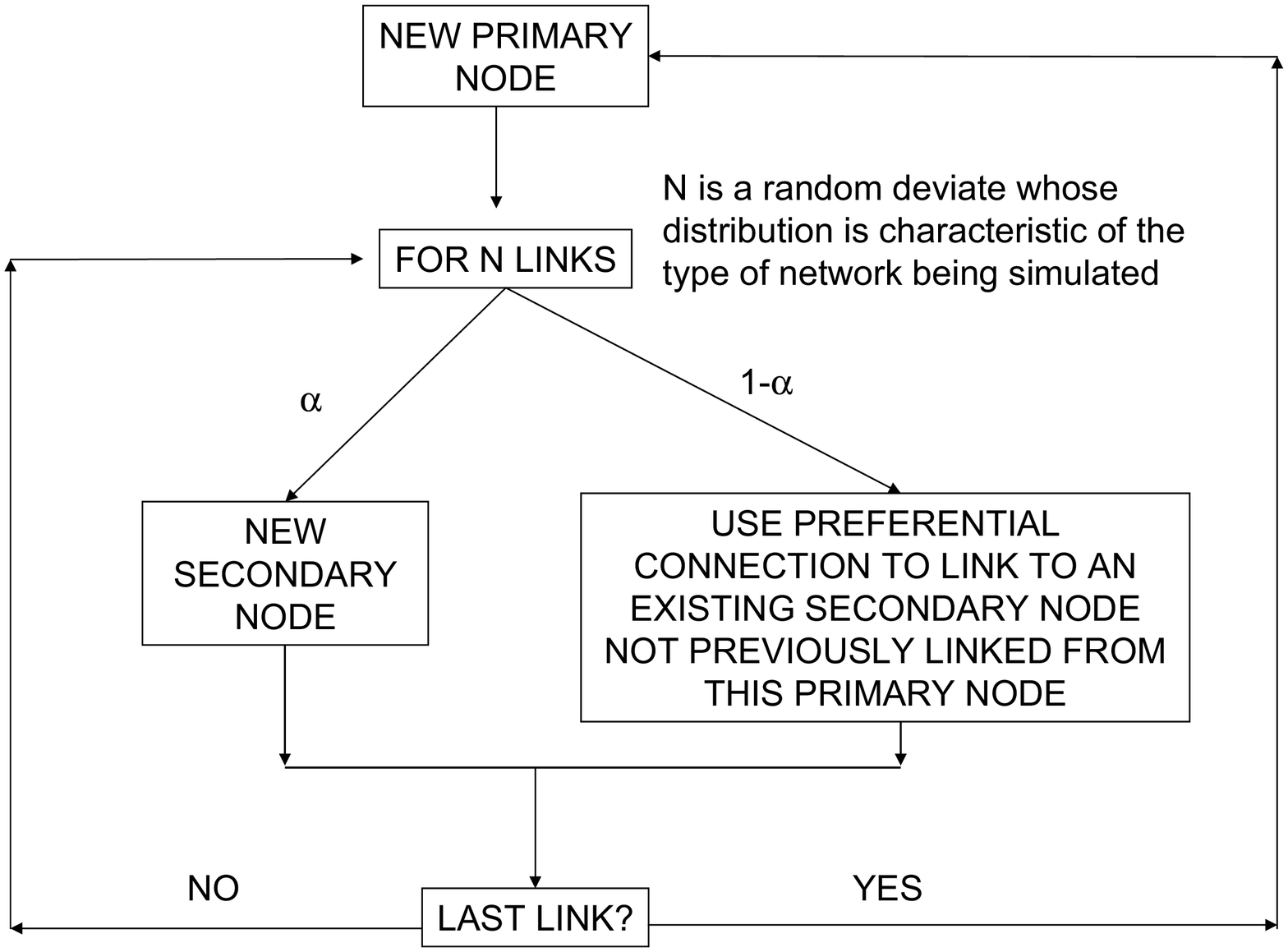}}%
\caption{Diagram of a basic bipartite Yule process.\label{basic}}
\end{figure}

\begin{itemize}
\item The network grows by adding primary nodes one at a time.

\item When a new primary node is added, it links to $N$ secondary nodes.
 $N$ is a random deviate drawn from a discrete probability distribution
that is a characteristic of the type of network being modeled. For
paper-reference networks $N$ is lognormally distributed
\cite{morris04a}, while for paper-author networks $N$ is 1-shifted
Poisson distributed \cite{goldstein04group} \cite{morris04b}. For
paper-journal networks, $N$ is unity, since a paper is only linked
to one journal, the one in which it was published. As defined here,
a primary entity does not link to any specific secondary entity more
than once.

\item For each of the $N$ links, there is a probability, $\alpha$, that it will link to a newly
appearing secondary node.

\item If a link happens to be to an existing secondary node, the linked node is selected using
preferential attachment, that is, the probability of linking to a
secondary node is proportional to the number of links that the node
possesses.
\end{itemize}

The stationary distribution of the link degree of the secondary
nodes is a Yule distribution \cite{johnson92}\cite{simon55}, a power
law whose exponent is $1+1/(1-\alpha)$. The stationary distribution
is independent of the distribution of $N$, but for finite
collections of papers the distribution of $N$ profoundly affects the
tail of the distribution \cite{morris04a}.

\section{Practical bipartite Yule processes}
In practice, the basic bipartite Yule process outlined in the
proceeding section must be modified to account for the
characteristics of the specific type of bipartite network being
studied.

\subsection{Paper-reference Yule process}
Figure 4 shows a diagram of a bipartite Yule process modified for
the characteristics of paper-reference networks. The details of this
model, its scope, and a discussion of evidence of the its validity,
appear in \cite{morris04a}. Paper-reference networks in collections
of papers covering scientific specialties are characterized by the
accretion of highly cited exemplar references, which are cited at
rates far higher than would be predicted by simple preferential
attachment. These exemplar references tend to appear during the
initial growth of the network and their rate of appearance decreases
exponentially as papers are added to the collection.

As each paper is added to the collection, it links to a lognormally
distributed number of references, as discussed in \cite{morris04a}.
For each reference cited by a paper, there is a probability $\alpha$
that the citation is to a newly appearing reference. When a new
reference appears, there is a small probability that the reference
will be a highly attractive exemplar reference. If so, the reference
receives a large initial attraction, $A_0$. Newly created
non-exemplar references received no initial attraction. If a
citation is to an existing reference, the probability that any
particular existing reference will be cited is proportional to the
sum of its attraction plus the number of times it has been cited. A
specific reference can not be cited more than once by a paper.

\begin{figure}
\resizebox{.45\textwidth}{!}{%
\includegraphics{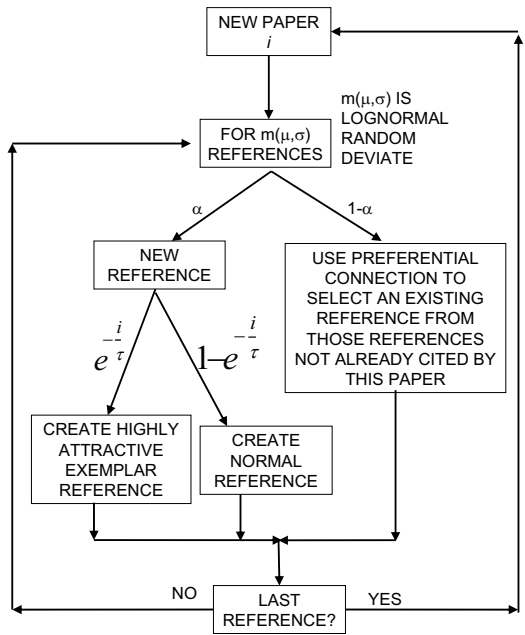}}%
\caption{Diagram showing a bipartite Yule process for
paper-reference networks.\label{prproc}}
\end{figure}

\subsection{Paper-author Yule process}
Figure 5 shows a diagram of the basic bipartite Yule process
modified for the characteristics of paper-author networks. The
details of this model, its scope, and a discussion of evidence of
the its validity, appear in \cite{goldstein04group} and
\cite{morris04b}. In this case the Yule process is applied to teams
of researchers rather than individual researchers. As each paper is
added, there is a probability   that the paper will be authored by a
new research team.  If so, a team of $N_G$ authors is added to the
network, but only $N(\lambda)$ appear as authors of the team's first
paper, where $N(\lambda)$ is a random deviate drawn from a 1-shifted
Poisson distribution whose parameter is $\lambda$.  If choosing an
existing team,  the teams are chosen using preferential attachment,
that is, the probability that a team will author the new paper is
proportional to the number of papers that the team has previously
published.

\begin{figure}
\resizebox{0.45\textwidth}{!}{%
\includegraphics{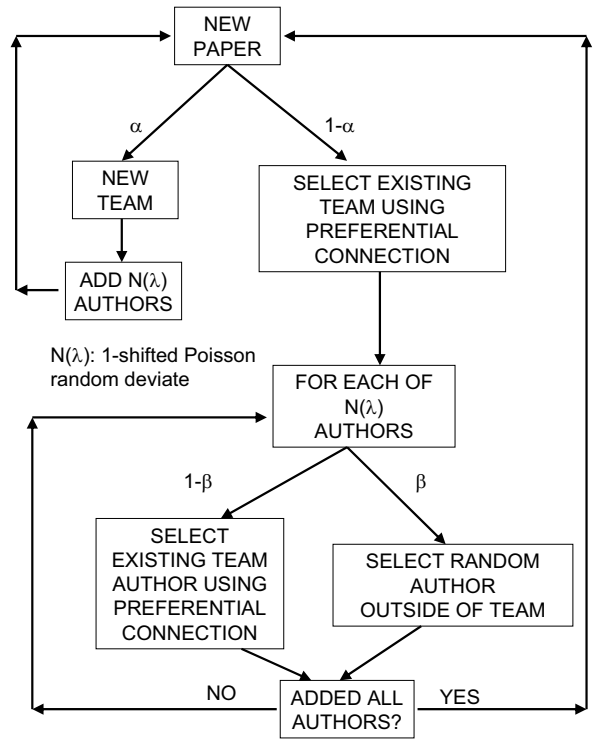}}%
\caption{Diagram showing a bipartite Yule process for paper-author
networks.\label{paproc}}
\end{figure}

When selecting authors for an existing team's paper, $N(\lambda)$
authors are chosen and the authors are selected using preferential
attachment, specifically, the probability of selecting an author is
proportional to 1 plus the number of papers that the author has
published. Inter-team collaborations (weak ties) are modeled as
random events; when an existing author is to be selected there is a
probability $\beta$ that the author will be drawn randomly from some
other team.

\section{Network metrics}
Simulation using a bipartite Yule process fully preserves the
topology of the network phenomenon being studied. The adjacency
matrix for a bipartite network is a roughly lower triangular
rectangular matrix. Figure 6 shows the adjacency matrices of the
paper-reference network, paper-author network, and paper-journal
network in an actual collection of papers.

\begin{figure*}
\resizebox{1\textwidth}{!}{%
\includegraphics{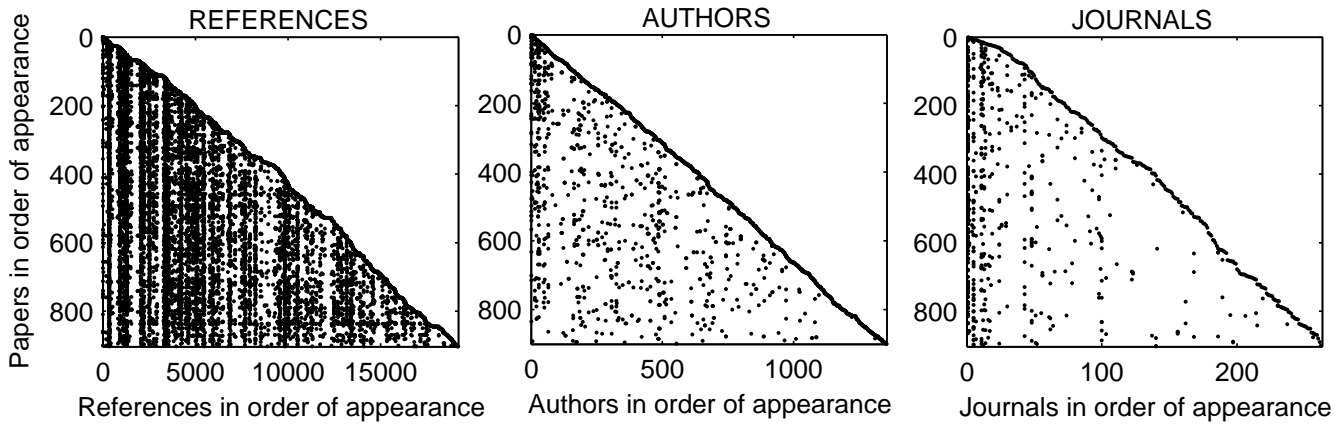}}%
\caption{Diagrams of adjacency matrices of bipartite networks in a
collection of 902 papers on the topic of complex
networks.\label{matrix}}
\end{figure*}

From each bipartite network, two co-occurrence networks can be
derived with their own characteristic topology.  For example, a
paper-reference network yields two unipartite networks, a
bibliographic coupling network of papers linked by common references
and a co-citation network of references linked by their common
papers. A paper-author network yields a  collaboration network of
authors connected by common papers and also a network of papers
connected by common authors.

Network metrics that characterize a bipartite network can be derived
from link degree distributions in the bipartite network and link
degree distributions in the associated unipartite co-occurrence
networks. Many of these metrics can be tied to indicators of the
underlying research process generating the collection of papers.

A set of useful metrics for paper-reference networks includes:
\begin{itemize}
\item \textit{reference per paper distribution} - This tends to be a
lognormal distribution whose mean, $m$, is from 15 to 30 references
per paper \cite{morris04a}.
\item \textit{paper per reference
distribution} - This tends to be a power-law distribution with a
characteristic exponent that ranges from 2 to 4
\cite{naranan71}\cite{redner98}.
\item \textit{bibliographic coupling strength per
paper pair distribution} - This is the link weight distribution of
the bibliographic coupling network.

\item \textit{co-citation coupling strength per reference pair distribution} -
This is the link weight distribution of the co-citation network.
\item \textit{bibliographic coupling clustering coefficient
distribution} - This the distribution of the clustering coefficients
for the bibliographic coupling network.
\end{itemize}
In paper-reference networks, the mean references per paper is
typically about 30, while the mean papers per reference is typically
about 1.4, the mean of a zeta (pure power-law) distribution with
exponent of 3. This constrains the ratio of references to papers in
the collection to be about 20, that is, a collection of papers
typically has about 20 times more references than papers.

A set of useful metrics for paper-author networks includes.
\begin{itemize}
\item \textit{authors per paper distribution} - This tends to be a 1-shifted
Poisson distribution whose mean varies from 2 for fields such as
mathematics to more than 10 for biomedical fields \cite{morris04b}.
\item \textit{paper per author distribution} - This tends to be a
power-law (Lotka's Law), whose exponent ranges from 2 to 4
\cite{lotka26}.
\item \textit{collaborating author distribution} - This is the
distribution of the number of unique co-authors per author in the
collection, and is the link degree distribution of the unweighted
co-authorship network.
\item \textit{co-authorship per author pair
distribution} - This is the link weight distribution of the weighted
co-authorship network.
\item \textit{co-authorship clustering coefficient
distribution} - This is the clustering coefficient of the unweighted
co-authorship network.
\item \textit{minimum co-authorship path length
distribution} - This is the distribution of minimum pathlengths
between author pairs in the unweighted co-authorship network.
\end{itemize}

\section{Examples}
\subsection{Example simulation of paper-reference network}
The Yule model for paper-reference networks was tested on a
collection of papers that cover the topic of complex networks.  This
collection was gathered on September 8th, 2003 from ISI's Web of
Science product using a series of queries to find all papers that
cite key references and authors in the specialty.  The collection
contains 902 papers with 31355 citations to 19185 references.  The
Yule parameter, $\alpha$, estimated by dividing the number of
references by the number of citations to references, is 0.61.  The
mean references per paper is 34.8. The parameters used for the
bipartite Yule simulation of this collection can be found in
\cite{morris04a}.

\begin{figure*}
\resizebox{.9\textwidth}{!}{%
\includegraphics{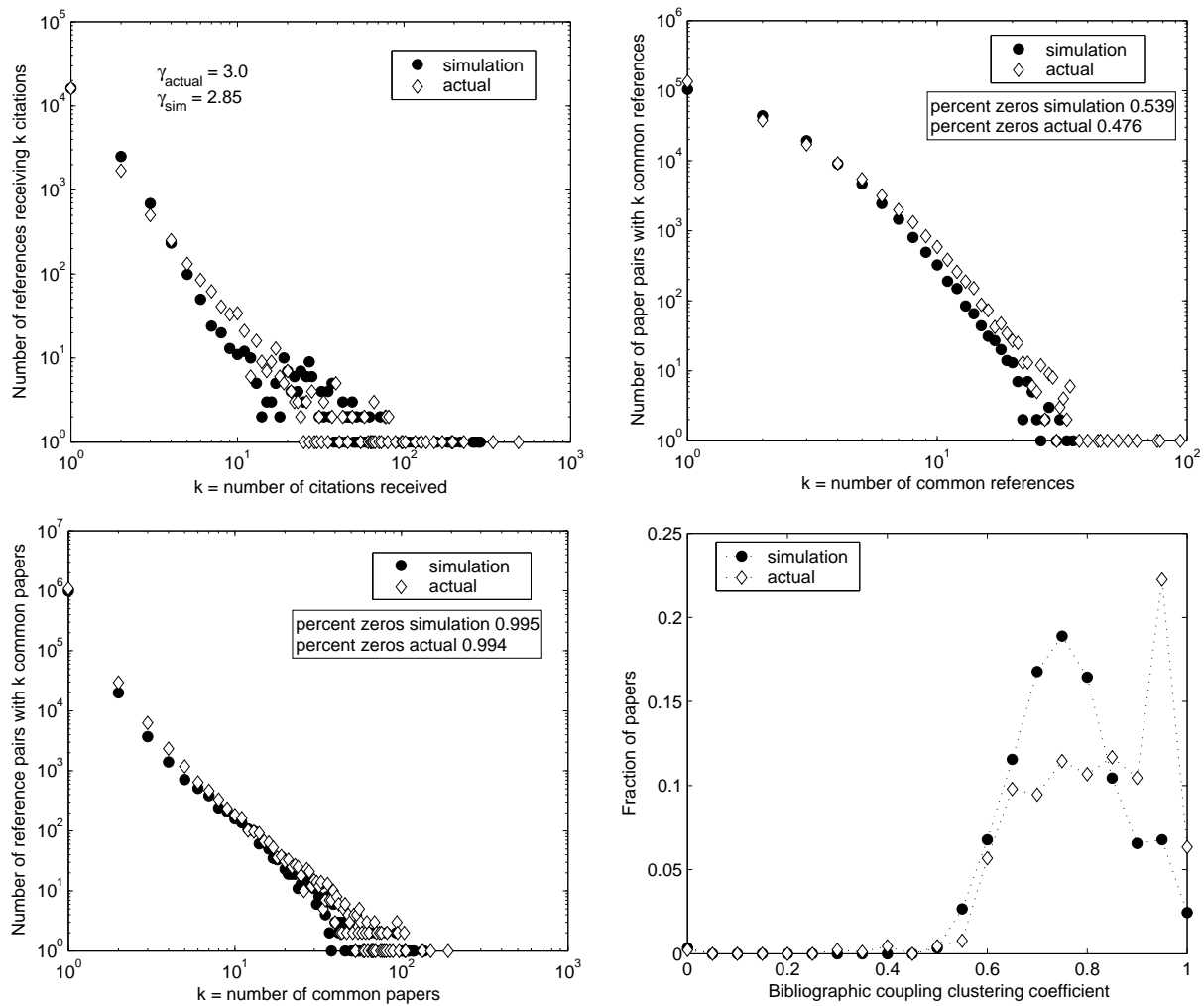}}%
\caption{Comparison plots of paper per reference frequency (upper
left), bibliographic coupling strength frequency (upper right),
co-citation strength frequency (lower left), and bibliographic
coupling clustering coefficient distribution (lower right), from a
collection of 902 papers on the topic of complex networks.
\label{pr_results}}
\end{figure*}

Figure 7 show plots comparing network metrics from the actual data
to a Yule simulation of network growth. The upper left plot is of
papers per reference frequencies. Maximum likelihood expectation
(MLE) estimated power-law exponents are 3.0 for the actual
frequencies, and 2.85 for the simulation. The paper-reference Yule
process mimics the phenomenon of exceptionally highly cited exemplar
references in the extreme lower right of the plot.  The upper right
plot is of frequency of bibliographic coupling strength per paper
pair. The Yule process-based simulation frequencies match the actual
frequencies well. The series of high bibliographic coupling strength
pairs in the lower right from actual data corresponds to pairs of
review papers with long lists of almost identical references, a
phenomenon not modeled by the Yule process. The lower left plot of
Figure 7 is of frequency of co-citation strength per reference pair.
The simulated frequencies match the actual frequencies well across
the whole plot. The lower right plot is of bibliographic coupling
clustering coefficient distribution. The simulated distribution
matches the shape and scale of the actual data.

\subsection{Example simulation of a paper-author network}
The Yule model for paper-author networks was tested on three
collections of papers representing specialties with a wide range of
collaboration intensities. A collection of 1391 papers on the topic
of distance learning with 51\% single-authored papers represents a
specialty with little collaboration. A collection of 900 papers on
the topic of complex networks with 21\% single-authored papers
represents a specialty with typical amount of collaboration.
Finally, a collection of 3095 papers on the topic of atrial ablation
with 7\% single-authored papers represents a specialty with heavy
collaboration \cite{morris04b}. The parameters used for bipartite
Yule simulation of these paper-author networks can be found in
\cite{morris04b}.

Figures 8, 9 and 10 show the comparison of Yule model simulations to
actual data for these three collections using two metrics: 1) paper
per author frequency (Lotka's Law), and 2) collaborating author
frequency.

\begin{figure*}
\resizebox{.9\textwidth}{!}{%
\includegraphics{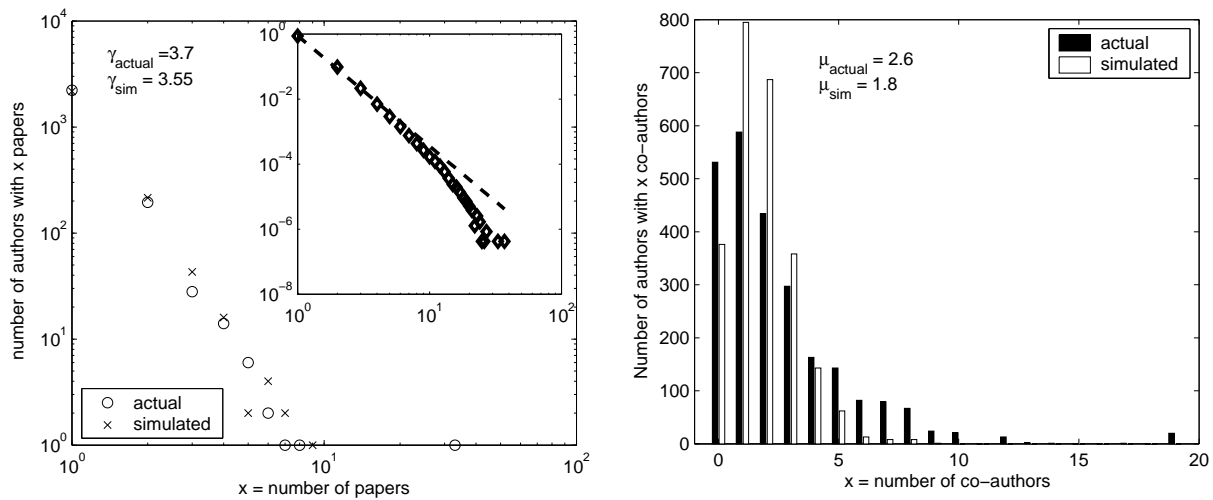}}%
\caption{Comparison of bipartite Yule simulation against actual data
 for plots of paper per author frequencies and collaborating author
 frequencies for the distance education paper collection.\label{distance}}
\end{figure*}

\begin{figure*}
\resizebox{.9\textwidth}{!}{%
\includegraphics{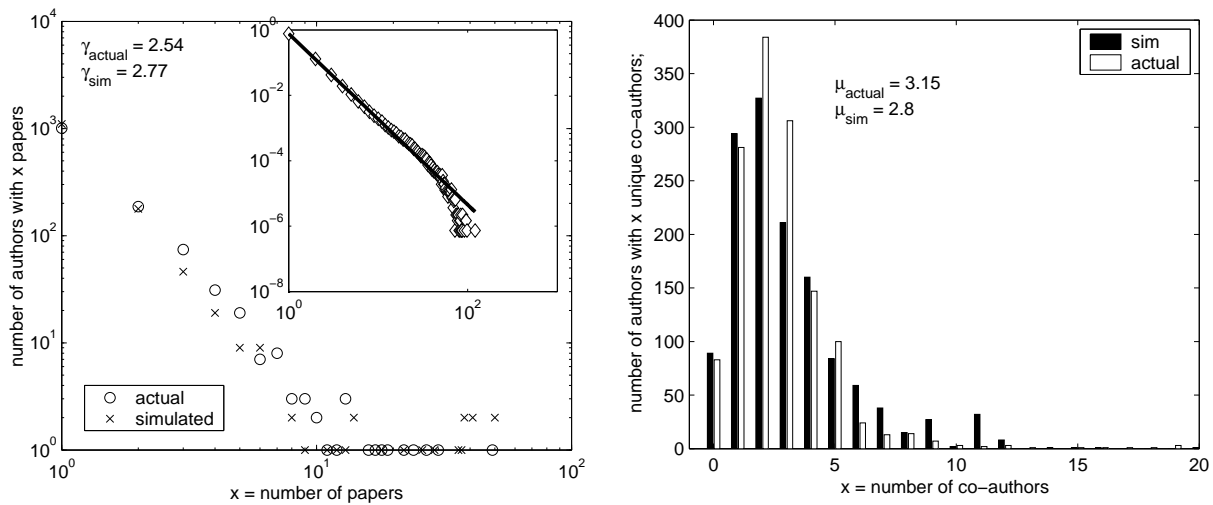}}%
\caption{Comparison of bipartite Yule simulation against actual data
 for plots of paper per author frequencies and collaborating author
 frequencies for the complex networks paper collection.\label{networks}}
\end{figure*}

\begin{figure*}
\resizebox{.9\textwidth}{!}{%
\includegraphics{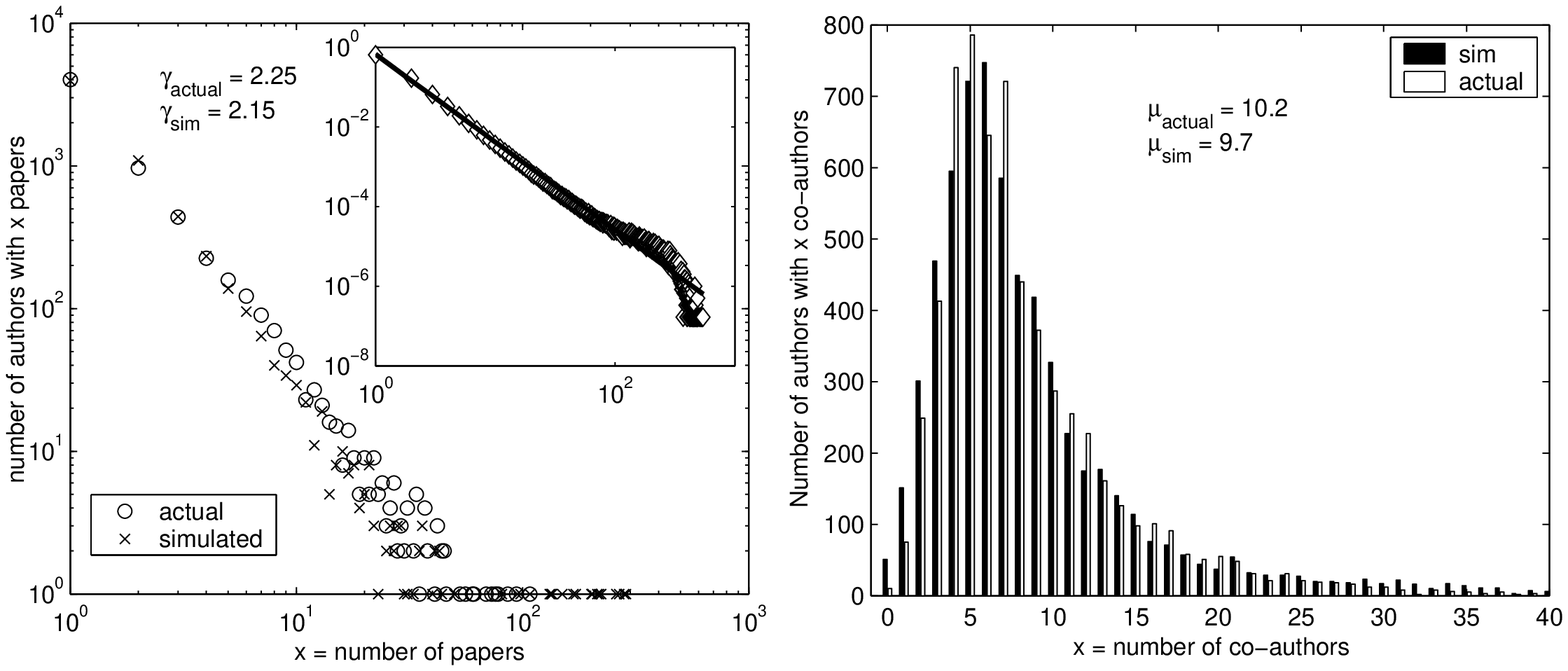}}%
\caption{Comparison of bipartite Yule simulation against actual data
 for plots of paper per author frequencies and collaborating author
 frequencies for the atrial ablation paper collection.\label{atrial}}
\end{figure*}

The left plots in Figures 8, 9 and 10 are paper per author frequency
plots. The bipartite Yule process produces excellent matches to
actual data. The inset plots show Yule model predicted paper per
author distributions derived by gathering statistics from 1000
simulations for each collection. A line representing an MLE fitted
zeta (pure power-law) distribution is shown in each inset. The Yule
model produces excellent fits to the zeta distribution for all three
collections, confirming the Yule model's usefulness as a predictor
of Lotka's Law. Note that the deviation of the distributions from
the zeta distribution in the tail of the distributions is due to
truncating the simulations at the number of papers in each
collection.  The plots on the right side of Figures 8, 9 and 10 show
that the bipartite Yule model produces good matches of collaborating
author frequencies to actual data across the wide rage of
collaboration intensities represented by the three collections.

\begin{figure}
\resizebox{.5\textwidth}{!}{%
\includegraphics{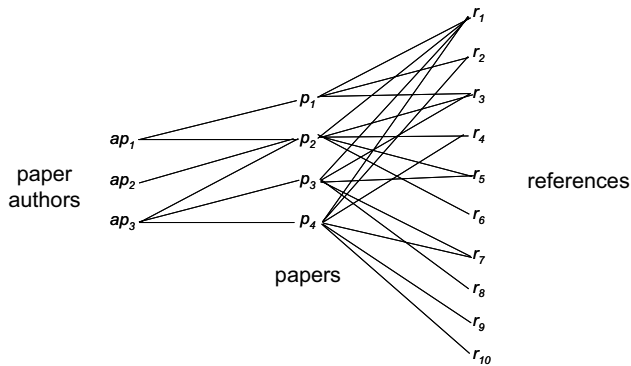}}%
\caption{Example of coupled bipartite networks. The paper-author
network is coupled to the paper-reference network through common
papers. \label{example}}
\end{figure}

\section{Future work}
The research on bipartite Yule processes discussed here will be
extended to modeling of coupled bipartite networks. Figure 10 shows
an example of coupled bipartite networks, where a paper-author
network is coupled to a paper reference network through common
papers. The challenge is to invent a model that reproduces the
correlation of groups of authors to groups of references, a
phenomenon that cannot be modeled using two separate bipartite
processes.

\bibliography{yule}

\end{document}